\documentclass[aps,10pt,pre,twocolumn,amsmath,amssymb,amsfonts]{revtex4-1}

\usepackage{graphicx}

\newcommand{\no}{\nonumber\\}
\renewcommand{\r}{\rangle\!\rangle}

\renewcommand{\>}{\rangle}
\newcommand{\<}{\langle}

\def\Eq#1{Eq.(\ref{#1})}
\def\Eqs#1{Eqs.(\ref{#1})}
\def\Fig#1{Fig.~\ref{#1}}
\def\Figs#1{Figs.~\ref{#1}}

\def\no{\nonumber\\}
\def\r{\rangle\!\rangle}

\def\>{\rangle}
\def\<{\langle}

\def\adg{a^\dagger}

\def\half{\tfrac{1}{2}}

\def\xh{\hat{x}}
\def\tr{\text{Tr}}
\def\dg{\dagger}
\def\lam{\lambda}

\def\w{\omega}
\def\wo{\omega_0}
\def\del{\delta}
\def\gam{\gamma}

\def\bt{\beta}
\def\eps{\epsilon}
\def\tt{\theta}
\def\Nt{\tilde{N}}
\def\rt{\tilde{r}}

\def\at{a}
\def\btil{b}
\def\ct{\tilde{c}}

\def\gt{\tilde{g}}
\def\htt{\tilde{h}}

\def\xt{\tilde{x}}

\def\rhot{\tilde{\rho}}

\def\nt{m}

\def\nuv{\vec{\nu}}

\def\L{\mathcal{L}}
\def\Lt{\tilde{L}}
\def\Kdt{\tilde{K}_d}
\def\nb{\bar{n}}

\def\fh{\hat{f}}
\def\d{\partial}

\def\eq{\text{eq}}
\def\ini{\text{ini}}
\def\ome{\omega}
\def\wk{\omega_k}

\begin{document}

\title{Reduced dynamics of two oscillators collectively coupled to a thermal bath}

\author{B.~A.~Tay}
\email{BuangAnn.Tay@nottingham.edu.my}
\affiliation{Foundation Studies, Faculty of Engineering, University of Nottingham Malaysia Campus, Jalan Broga, 43500 Semenyih, Selangor, Malaysia}

\date{\today}

\begin{abstract}

We study the reduced dynamics of a pair of non-degenerate oscillators coupled collectively to a thermal bath. The model is related to the trilinear boson model where the idler mode is promoted to a field. Due to nonlinear coupling, the Markovian master equation for the pair of oscillators admits non-Gaussian equilibrium states, where the modes distribute according to the Bose-Einstein statistics. These states are metastable before the nonlinear coupling is taken over by linear coupling between the individual oscillators and the field. The Gibbs state for the individual modes lies in the subspace with infinite occupation quantum number. We present the time evolution of a few states to illustrate the behaviors of the system.

\end{abstract}

%\pacs{03.65.Yz;03.65.Ge;05.70.Ln}
%\keywords{Markovian master equation}

\maketitle

%%%%%%%%%%%%%%%%%%%%%%%%%%
%       Sectionn         %
%%%%%%%%%%%%%%%%%%%%%%%%%%
\section{Introduction}

We consider a composite system of two non-degenerate oscillators coupled collectively to a bath, i.e.~the change in the occupation quantum number of one oscillator mode due to the environmental influence induces a corresponding change in the other mode. In the studies of environmental influence on a pair of oscillators, the oscillators are usually coupled separately to the bath through linear interactions \cite{Dodd04,*Liu07,*Hu08,*Paz08,*Xiang09,*Galve10,*Isar11}. In contrast, our main focus is on the three-body interactions between the oscillators and the field modes of the bath.

The system is related to the trilinear boson model in quantum optical systems, where it is used to describe the process of parametric amplification and frequency conversion \cite{Louisell61,Glauber67,Walls69,Walls70}. The two oscillators then play the roles of the pump mode and the signal mode, respectively, which are coupled to the idler or vibrational mode of a nonlinear medium in which nonlinear interactions are assumed to be dominant. By promoting the idler mode to a field and assuming that the coupling is weak, we can employ the standard open quantum system approach \cite{Gardiner,Breuer} to study the damping of the system of oscillators in a thermal bath of the field.

As far as we know the reduced dynamics of this model has not been discussed before for the general sectors in the non-integrable region. We find that the Markovian master equation as a time independent eigenvalue problem can be solved analytically. Due to the collective coupling between the oscillators and the bath, the reduced dynamics exhibits the behaviors of finite-level systems \cite{Nielsen,Alicki07,Hioe82} under the cascade process, even though we are dealing with a continuous variable system.

The Hamiltonian of the system has the same formal structure as the Lee model for the bosonic systems \cite{Lee,Sudarshan92}, and the one-particle sector of its oscillators subsystem is equivalent to the Friedrichs model \cite{Friedrichs48}. The Friedrichs-Lee model was originally devised to study the effect of perturbation on the spectra in the Hilbert space \cite{Friedrichs48}, the mathematical structure of renormalizable quantum field theory \cite{Lee,Kallen55}, and later on to study the non-integrable systems where resonance states emerge \cite{Glaser56,Petrosky91,Antoniou98}.

The reduced dynamics for this type of interaction shows a few unique features. It gives rise to a family of non-Gaussian equilibrium states confined to their respective irreducible subspaces, whereas the Gibbs state of the individual oscillators is recovered in the subspace with unrestricted occupation quantum number. The oscillator modes in the equilibrium eigenstates distribute according to the Bose-Einstein statistics \cite{Reif}. These states are metastable before the nonlinear coupling is taken over by linear coupling between the individual oscillators and the field.

In our discussion, we first present the Hamiltonian of the system in Sec.~\ref{sec-model}. The Markovian master equation of the reduced system and its bosonic representation are then presented in Sec.~\ref{sec-KEsu2}. We then solve for the equilibrium states in Sec.~\ref{sec-EigProb}, and study the time evolution of some states in Sec.~\ref{sec-timeevo}.

%%%%%%%%%%%%%%%%%%%%%%%%%%
%       Sectionn         %
%%%%%%%%%%%%%%%%%%%%%%%%%%
\section{The Hamiltonian}
\label{sec-model}

We consider a system of two oscillators and a field in one dimensional space, labeled by $1, 2$, and $k$, respectively. The free Hamiltonian is
%%%
\begin{align}   \label{H0}
    H_0 &= \w_1 \adg_1 a_1 + \w_2 \adg_2 a_2 + \sum_k \wk \adg_k a_k \,,
\end{align}
%%%
where we use the units $\hbar=c=1$, and $\w_1, \w_2$ are the natural frequencies of the respective oscillators.  We assume that $\w_1>\w_2$, and the field is massless, with the dispersion relation $\ome_k=|k|$. The creation and annihilation operators $\adg_i$ and $a_i$ obey the commutation relations, $[a_i, \adg_j]=\del_{i,j}, i,j=1,2,k$. We normalize the field in a box with length $\Omega$, so that $k=2\pi i/\Omega$, with $i=1,2,3,\cdots$. The limit $\Omega\rightarrow \infty$ will be taken eventually, but we continue to use the discrete notation in the expressions below.

We assume that the oscillators are coupled collectively to the field through the interaction
%%%
\begin{align}   \label{HV}
    V = \lam \sum_k \frac{v(\wk)}{\sqrt{\Omega/2\pi}} (L_+ a_k+L_- \adg_k ) \,,
\end{align}
%%%
in analogy to the linear coupling model between an oscillator (labeled by $a,\adg$) and the field, i.e., $\adg a_k+a \adg_k$. A derivation of this interaction can be found in Ref.~\onlinecite{Walls69} in the context of trilinear boson model, where it is used to describe parametric amplification and frequency conversion in quantum optical systems \cite{Louisell61,Glauber67,Walls69,Walls70}. The $L_\pm$ are the ladder operators of the SU(2) algebra,
%%%
\begin{align}   \label{L+-}
        L_+ &= \adg_1 a_2 \,, &\qquad   L_-&= a_1 \adg_2 \,.
\end{align}
%%%
They raise and lower the $r$-quantum number of the composite system, respectively, see \Eqs{stn1n2}-\eqref{LNr-} below. In \Eq{HV}, $\lam$ is a dimensionless coupling constant, and $v(\wk)$ is a form factor that contains a high frequency cut-off to regularize the interactions.

The total Hamiltonian of the system, $H=H_0+V$, when written explicitly in terms of the individual mode, has a form similar to the Lee model for bosonic system \cite{Lee,Sudarshan92}. Its one particle sector of the oscillators subsystem is equivalent to the Friedrichs model \cite{Friedrichs48}.

The system possesses two independent constants of motion,
%%%
\begin{align}   \label{N}
        N&=\adg_1 a_1 + \adg_2 a_2\,, &\quad
        N_{\!1k}&=\adg_1 a_1 + \sum_k \adg_k a_k\,.
\end{align}
%%%
$N$ remains a constant of motion of the reduced dynamics when the field modes are traced out. In the unstable regime, the system develops a resonance at the frequency
%%%
\begin{align}  \label{w0}
    \wo&= \w_1-\w_2\,,
\end{align}
%%%
where the $1$-oscillator turns unstable and decays into the $2$-oscillator and the field \cite{Glaser56}. Complex poles corresponding to the unstable oscillator then arise in the complex energy plane.

From another point of view, the $a_i$s can be regarded as the normal modes of two degenerate oscillators coupled through the SU(2) coupling interactions \cite{SU2Chen11}. The system Hamiltonian $H$ is then unitarily related to a family of Hamiltonians by SU(2) transformation. The details are presented in App.~\ref{sec-SU2}. Therefore, the reduced dynamics discussed below is also applicable to these systems. It is then interesting to note that experimentally \cite{SU2AnisoOscLin12} it had been shown that spatial wave patterns generated by three-dimensional coherent waves obtained through the longitudinal and transverse coupling of laser modes in a cavity \cite{Chen06} are related to the eigenstates of the system with SU(2) coupling interactions. It may then be possible to realize the spatial profiles of a mixture of eigenstates of the system in the future.

We could gain further insights into the conditions under which the model is applicable by comparing the Hamiltonian to anharmonic interactions \cite{Prigogine62}. However, since this comparison is outside the main line of our discussion, we present it in App.~\ref{sec-anharmonic}.

%%%%%%%%%%%%%%%%%%%%%%%%%%
%       Sectionn         %
%%%%%%%%%%%%%%%%%%%%%%%%%%
\section{Markovian Master equation -- bosonic representation of SU(2)}
\label{sec-KEsu2}

%%%%%%%%%%%%%%%%%%%%%%%%%%
%       Sectionn         %
%%%%%%%%%%%%%%%%%%%%%%%%%%
\subsection{Markovian master equation}
\label{sec-ME}

The reduced dynamics of the two oscillators subsystem immersed in a thermal bath can be derived by tracing out the field degrees of freedom with standard methods \cite{Gardiner,Breuer}, or through the complex spectral representation \cite{Petrosky97,*Barsegov02}. In the derivation, we assume that the oscillators and the fields are initially factorizable, and we use the weak coupling limit, or equivalently, the $\lam^2t$-approximation \cite{VanHove57,Prigogine62} or the Born-Markov approximation \cite{Louisell72,Gardiner,Breuer}.

Since the structure of the interaction Hamiltonian is similar to the linear coupling model between a single oscillator and a field \cite{Breuer}, the master equation acquires the Kossakowski-Lindblad (KL) form \cite{Lindblad,Kossa76}, with a modified unitary part. The reduced dynamics is therefore completely positive too \cite{Lindblad,Kossa76}. The reduced density operator of the two oscillators, $\fh$, evolves according to the equation $\d\fh/\d t=-K\fh$, where
%%%
\begin{align}   \label{K}
    K=K_0+K_d
\end{align}
%%%
can be decomposed into a unitary part,
%%%
%\begin{widetext}
\begin{align}   \label{K0}
K_0&\fh
    = i[H_0',\fh]\,,\\
    H_0'&=(\w_1-\del\w_1)\adg_1 a_1+(\w_2-\del\w_2)\adg_2 a_2-\del\ome'_0 \adg_1 a_1\adg_2 a_2 \no
    &= (\wo -\del\wo) L_0-\tfrac{1}{2} (\ome'_0-\del\ome'_0)N +\del\ome'_0(L_0^2-\tfrac{1}{4}N^2)\,, \label{H0'}
%    &=\tfrac{i}{2} (\ome'_0-\del\ome'_0)[N, \fh] +i(\wo -\del\wo) [L_0, \fh]\no
%    &\qquad+i \del\ome'_0[L_0^2-\tfrac{1}{4}N^2, \fh]\,,
\end{align}
%\end{widetext}
%%%
and a dissipative part,
%%%
\begin{align}  \label{Kd}
    &K_d\fh=-\half  \gam  \nb_0 (2 L_+ \fh L_- - L_-L_+ \fh -\fh L_-L_+ ) \no
            & -\half \gam (\nb_0+1) (2 L_- \fh L_+ - L_+L_- \fh - \fh L_+L_-)   \,.
\end{align}
%%%
In the first equality of \Eq{H0'}, we have presented $H'_0$ in terms of the creation and annihilation operators to exhibit the frequency renormalization $\del \w_i$ to the individual oscillator. The explicit expressions of the coefficients are
%%%
\begin{subequations}
\begin{align} \label{delw0'}
            \del\ome'_0&\equiv\del\w_1+\del\w_2\,, \\
    \del\wo&\equiv\del\w_1-\del\w_2\,,\label{delw0}\\
%        \wo&\equiv \w_1-\w_2\,, \qquad\qquad
        \del\w_1 &\equiv \frac{\lam^2}{\Omega/2\pi}\sum_k {\rm P} \frac{|v(\wk)|^2}{\wk-\wo}(\nb_k+1)\,,\label{delmu1}\\
     \del\w_2 &\equiv -\frac{\lam^2}{\Omega/2\pi}\sum_k {\rm P} \frac{|v(\wk)|^2}{\wk-\wo}\nb_k\,,\label{delmu2}\\
     %\label{nk}
   \gam &= 2\pi \lam^2|v(\wo)|^2 \,, \label{gam0}
\end{align}
\end{subequations}
%%%
where $\wo$ is the resonant frequency \eqref{w0} and $\gam$ is the decay constant. We note that the natural frequencies of the oscillators are renormalized with opposite signs, compare \Eqs{delmu1} with \eqref{delmu2}. In scattering problems, the average number of field modes is zero $\nb_k=0$. In this case, we find that only the frequency of the $1$-oscillator is renormalized, consistent with the discussion in Ref.~\onlinecite{Lee}.

We assume that the field modes are in thermal equilibrium satisfying the Bose-Einstein distribution $\nb_k= 1/[\exp(\wk \bt)-1]$, where $\bt=1/(k_B T)$. For the resonant mode, we label its occupation number by
%%%
\begin{align}
     \nb_0\equiv \frac{1}{e^{\wo \bt}-1} \label{nb0}\,.
\end{align}
%%%

%%%%%%%%%%%%%%%%%%%%%%%%%%
%       Sectionn         %
%%%%%%%%%%%%%%%%%%%%%%%%%%
\subsection{SU(2) bosonic representation}
\label{sec-su2boson}

The generators of the SU(2) group in terms of the bosonic representation \cite{Schwinger} are
%%%
\begin{subequations}
\begin{align}   \label{L0}
    L_0 &=  (\adg_1 a_1-\adg_2 a_2)/2\,,\\
        L_1 &=  (L_+ +L_- )/2\,, &\,
        L_2&= (L_+-L_- )/2i \,.\label{L12}
\end{align}
\end{subequations}
%%%
They obey the commutation relations $[L_i,L_j]=i\eps_{ijk} L_k$. The Casimir operator of the SU(2) is \cite{Schiff}
%%%
\begin{align}   \label{Cas}
    L^2&\equiv \half(L_+L_-+L_-L_+)+L_0^2 =L_+L_-+L_0(L_0-1)\,,
\end{align}
%%%
which commutes with the $L_i$s, $[L^2,L_i]=0$. The total occupation number of both oscillators as denoted by $N$ \eqref{N} remains a constant of motion of the reduced dynamics. It commutes with the generator of the SU(2) group, $[N, L_i]=0, i=0, \pm$. The quantum number $N$ will be used to label the irreducible representation, see \Eq{stn1n2} below.

\begin{figure}[tb]
\centering
\includegraphics[width=3.4in]{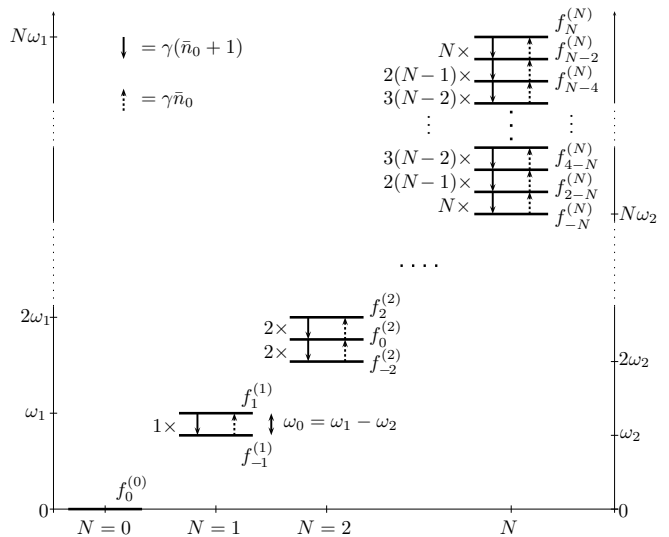}
\caption{Energy level diagram and the rate of transitions between different levels in the $f^{(N)}$ subspace (between diagonal elements), see \Eq{fdiag} for the definition of $f^{(N)}_r$. The transition rates between two levels are proportional to the numerical constants on the left of the levels. Transitions between different irreducible $f^{(N)}$ subspaces are forbidden by the SU(2) symmetry of the reduced dynamics.}
\label{fig1}
\end{figure}

We make use of the occupation number basis
%%%
\begin{align}   \label{|nn>}
        |n_1,n_2\>=\frac{(\adg_1)^{n_1}}{\sqrt{n_1 !}}
                    \frac{(\adg_2)^{n_2}}{\sqrt{n_2 !}}|0,0\>\,,
\end{align}
%%%
and denote a state in the irreducible subspace labeled by $N$ as
%%%
\begin{align}   \label{stn1n2}
    |r\>_N \equiv |n_1, n_2\>  \,,
\end{align}
%%%
where
%%%
\begin{align}   \label{rN}
    r&\equiv n_1-n_2\,, &\qquad N&\equiv n_1+n_2\,,
\end{align}
%%%
are related to the eigenvalues of $L_0$ and $L^2$ through \Eqs{L0st} and \eqref{L2st} below, respectively. Using these labels we can establish the following relations,
%%%
\begin{subequations}
\begin{align}   \label{LNr+}
    L_+ |r\>_N %&= \adg_1 a_2 |n_1, n_2 \>
                &= \half \sqrt{( N+r+2)( N-r)} |r+2\>_N \,,\\
    L_- |r\>_N %&= a_1 \adg_2 |n_1, n_2 \>
                &=\half\sqrt{( N+r)( N-r+2)} |r-2\>_N \,,\label{LNr-}\\
    L_0 |r\>_N &= \half r |r\>_N \,,\label{L0st}\\
    L^2|r\>_N&= \tfrac{1}{4} N(N+2) |r\>_N \,,\label{L2st}\\
    N |r\>_N &= N |r\>_N \,.\label{NrNr}
\end{align}
\end{subequations}
%%%
Whenever a 1-oscillator is created, a 2-oscillator is annihilated, and vice versa. Consequently, the index $r$ changes in step of $\pm 2$ under $L_\pm$. There is a total number of $N+1$ substates in each irreducible subspace, and $r$ ranges from $-N, -N+2, \ldots, N-2, N$. The highest and lowest states are $|N\>_N$ and $|\!-\!\!N\>_N$, annihilated by the raising and lowering operators, $L_\pm|\!\pm \!N\>_N=0$, respectively. The state $|r\>_N$ has energy
%%%
\begin{align} \label{ErN}
        E&=\w_1 n_1+\w_2 n_2 = \half (N \ome'_0 +r \wo ) \,,
\end{align}
%%%
see \Fig{fig1} for a plot of the energy levels.
%%%
%%%
They are non-degenerate if the ratio $\w_1/\w_2$ is not a rational number.

We denote the basis in the Liouville space by
%%%
\begin{align}   \label{str}
    f^{(N, \Nt)}_{r;\rt} \equiv |r\>_N {_{\Nt}}\< \rt| = |n_1,n_2\>\<\nt_1,\nt_2|  \,.
\end{align}
%%%
We find that using this notation is more convenient for our later discussion since it is more compact and it manifests the fact that $N$ and $\Nt$ are constants of motion of the reduced dynamics. Since the $L_\pm$ operators come in pairs in $K_d$ \eqref{Kd}, $K f^{(N, \Nt)}_{r;\rt}$ is a linear combination of $f^{(N, \Nt)}_{r;\rt}, f^{(N, \Nt)}_{r\pm 2;\rt\pm 2}$. Consequently, the quantity
%%%
\begin{align} \label{Delr}
  \nu \equiv r-\rt
\end{align}
%%%
is a constant of motion under $K$, and the basis states in each $f^{(N,\Nt)}$ subspace are connected to the others with the same $\nu$ value only, under the reduced dynamics.

It is interesting to note that the generator of the time evolution $K$ is invariant under a rotation along the $L_0$ axis, as shown in App.~\ref{sec-L0sym}. Furthermore, the SU(2) generalized coherent states \cite{Perelomov72,Wodkiewicz85,Buzek89} reside in the $f^{(N,N)}$ subspace, see App.~\ref{sec-SU2cohst} for details.

%%%%%%%%%%%%%%%%%%%%%%%%%%
%       Sectionn         %
%%%%%%%%%%%%%%%%%%%%%%%%%%
\section{Equilibrium states}
\label{sec-EigProb}

In this section, we obtain the set of equilibrium states for the irreducible subspaces. These states are non-Gaussian states in the coordinate space, as opposed to the Gaussian Gibbs state for an oscillator in a thermal bath. We will also show that the oscillator modes in these equilibrium states distribute according to the Bose-Einstein statistics, and the Gibbs states of the individual oscillator are recovered when the occupation quantum numbers for these modes are not restricted.

We first note that the equilibrium states have non-zero trace and can be decomposed into the diagonal basis state labeled by
%%%
\begin{align} \label{fdiag}
f^{(N)}_r\equiv f^{(N, N)}_{r;r}\,,
\end{align}
%%%
whenever $\Nt=N$ and $\rt=r$. We can then write the equilibrium state as a sum of the diagonal basis states,
%%%
\begin{align}   \label{Neigenv}
        f^{(N)}_\eq &= \frac{1}{Z_N} \sum_{n=0}^N p_{N-2n} f^{(N)}_{N-2n} \,,
%        \frac{1}{Z_N} \bigg[ p_N f^{(N)}_N +p_{N-2 }f^{(N)}_{N-2} + \ldots \no
%            &\qquad\qquad  +p_{-N+2} f^{(N)}_{-N+2} + p_{-N} f^{(N)}_{-N}\bigg]\,,
\end{align}
%%%
where $Z_N$ is a normalization constant, and $p_{N-2n}$ are coefficients to be obtained below. Note that since each $f^{(N)}_r =|n_1\>\<n_1| \otimes|n_2\>\<n_2|$ is separable,  $f^{(N)}_\eq$ is a separable state.

The action of $K$ on these states is
%%%
\begin{align}   \label{KABC}
        K f^{(N)}_r&= u_r f^{(N)}_{r+2}+ v_r f^{(N)}_r+w_r f^{(N)}_{r-2}\,,
\end{align}
%%%
where
%%%
\begin{subequations}
\begin{align}   \label{ABC}
%         a_r &\equiv -\frac{\gam}{4} \nb_0(N-r) (N+r+2) \,,\\
%          b_r &\equiv \frac{\gam}{2}\nb_0 [(N+1)^2 -r^2-1]+ \frac{\gam}{4}(N+r)( N-r +2) \,,\\
%          c_r &\equiv -\frac{\gam}{4}(\nb_0+1) (N+r) ( N-r+2) \,,
         u_r &= -\frac{\gam}{4} \nb_0\big[ (N+1)^2-(r+1)^2\big] \,,\\
          v_r &= \frac{\gam}{4}(2\nb_0+1) \big[(N+1)^2 -r^2-1\big]+ \frac{\gam}{2}r\,,\\
          w_r &= -\frac{\gam}{4}(\nb_0+1) \big[ (N+1)^2-(r-1)^2\big] \,,\label{ABCc}
\end{align}
\end{subequations}
%%%
by using \Eqs{LNr+}-\eqref{NrNr}. The transition rates between different energy levels within the same $f^{(N)}$ subspace are summarized in \Fig{fig1}. They can be read off from \Eqs{KABC}-\eqref{ABCc} directly to give
 %%%
\begin{subequations}
\begin{align}
    f^{(N)}_r &\rightarrow  f^{(N)}_{r+2} = \left\{ \begin{array}{ll}
                                                0\,, & \text{if} \quad r=N\,,\\
                                                u_r\,,& \text{if} \quad r\neq N\,,\\
                                                \end{array} \right. \\
              &\rightarrow  f^{(N)}_{r} =  v_r\,, \quad\,\,\, \text{all} \quad r \\
              &\rightarrow  f^{(N)}_{r-2} = \left\{ \begin{array}{ll}
                                                w_r\,, & \text{if} \quad r\neq -N\,,\\
                                                0 &\text{if} \quad r=-N\,.\\
                                                \end{array} \right.
\end{align}
\end{subequations}
The maximum rate is proportional to the numerical factor $N(N+2)/4$ for a transition between the $f^{(N)}_{\pm2}$ and the $f^{(N)}_0$ levels for even $N$, whereas the maximum rate is proportional to $(N+1)^2/4$ for a transition between the $f^{(N)}_1$ and the $f^{(N)}_{-1}$ levels for odd $N$ \cite{Breuer}.

Since the transitions between the energy levels in the $f^{(N)}$ subspace are very similar to the Dicke model for $N$ two-level systems collectively coupled to a field \cite{Dicke54,Gross82}, superradiance \cite{Dicke54} may also arise in this model. However, the energy levels for the two models are different, compare \Fig{fig1} in this paper to Fig.~1 in Ref.~\onlinecite{Dicke54}. This is because a symmetrization of the underlying states is required for $N$ two-level systems but not for a pair of non-degenerate oscillators.

The equilibrium condition, $K f^{(N)}_\eq=0$, can be written as a matrix equation,
%%%
\begin{align}   \label{Kfeq}
    \textbf{K}\cdot \textbf{f}^{(N)}_\eq =0\,,
\end{align}
%%%
where
%%%
\begin{align} \label{Kuvw}
&\textbf{K}=\no
&\left(\!\!\begin{array}{ccccccccc}
                v_N & u_{N\!-2} &0 & 0&\\
                w_N & v_{N\!-2} & u_{N\!-4} &0& && \ldots &\\
                0 & w_{N\!-2} & v_{N\!-4} & u_{N\!-6} &\\
                & & &&\!\ddots\! & & \\
                & && & &w_{6-\!N}&v_{4-\!N} & u_{2-\!N} & 0 \\
                & &\vdots& & &0&w_{4-\!N} & v_{2-\!N} & u_{-\!N} \\
                & && & &0 &0& w_{2-\!N} & v_{-\!N} \\
                \end{array}\!\!\right)
\end{align}
%%%
is a tridiagonal square matrix of dimension $N+1$, and the eigenvector is a column matrix,
%%%
\begin{align}   \label{amatrix}
    \textbf{f}^{(N)}_\eq = \frac{1}{Z_N} \left(\begin{array}{c}
                p_N\\
                p_{N-2}\\
                p_{N-4}\\
                \vdots\\
                p_{4-N}\\
                p_{2-N}\\
                p_{-N}
                \end{array}\right)\,.
\end{align}
%%%
The solution to the coefficients is
%%%
\begin{align}   \label{aN}
    p_{N-2n_2} &= \nb_0^{N-n_2}(1+\nb_0)^{n_2} \,, \qquad n_2=0, 1, 2, \cdots, N \,,
\end{align}
%%%
which can be checked to satisfy \Eq{Kfeq} easily. By requiring $\tr(f^{(N)}_\eq)=1$, we obtain the normalization constant
%%%
%\begin{subequations}
\begin{align}   \label{Zsuma}
    Z_N&\equiv \sum_{n_2=0, 1, 2, \cdots}^N p_{N-2n_2}=(1+\nb_0)^{N+1} -\nb_0^{N+1}\,,
\end{align}
%\end{subequations}
%%%
by using the relation $(1-r)(1+r+r^2+\cdots+r^N)=1-r^{N+1}$.

Furthermore, the equilibrium occupation probability $p_{2n_1-N}$ can be cast into the form (for fixed $N$),
%%%
\begin{subequations}
\begin{align} \label{canonicala}
        p_{N-2n_2}&=\left[e^{\bt\w_2}(1+\nb_0)\right]^N\, e^{-\bt E}\\
        &\propto e^{-n_1\bt\w_1}e^{-n_2\bt\w_2}\,,\label{canonicalb}
\end{align}
\end{subequations}
%%%
where $E$ is the energy of the configuration \eqref{ErN}. \Eq{canonicala} shows that $p_{N-2n_2}$ satisfies the canonical distribution for fixed $N$, whereas \Eq{canonicalb} shows that the system obeys the Bose-Einstein statistics with two energy modes \cite{Reif}, where the 1- and 2-oscillator states play the roles similar to the excited and ground states in a two-level system, respectively.

A special case occurs at zero $T$, or equivalently, when the field mode has zero occupation number, $\nb_0=0$. Damping to the system can then be caused only by the spontaneous emission of a field quantum \cite{Glauber67}, accompanied by the lowering and raising of the 1- and 2-oscillator's occupation quantum numbers, respectively. In this situation, the lowest state in each subspace, such as $f^{(N)}_{-\!N}$ in the diagonal subspaces, and $f^{(N,\Nt)}_{-\!N;-\!\Nt}$ for $N\neq \Nt$ in the off-diagonal subspaces, are annihilated by $K_d$. Hence, the subspace spanned by these states does not experience decoherence. Therefore, they should be included in the expression of the equilibrium state for $T=0$,
%%%
\begin{align}   \label{decohfree}
        f'_\eq&=\sum_{N=0,1,2,\cdots} c_N f^{(N)}_{-N} \no
        &\quad + \sum_{N,\Nt=0,1,2,\cdots}^{N\neq \Nt} \left( c'_{N\Nt} f^{(N,\Nt)}_{-\!N,\!-\!\Nt} +\text{h.c.}\right) \,,
\end{align}
%%%
where $c_N$ are real coefficients subjected to the normalization condition $\sum_N c_N=1$, and $c'_{N\Nt}$ are complex coefficients constrained by the positivity condition of $f'_\eq$. $f'_\eq$ is a separable state by inspection. Indeed, since
%%%
\begin{align}
    f^{(N,\Nt)}_{-\!N,-\!\Nt}=|0,\!N\>\<0,\!\Nt|=|0\>\<0|\otimes |N\>\<\Nt|\,,
\end{align}
%%%
we can factor out the overall 1-oscillator state $|0\>\<0|$ from the right hand side of \Eq{decohfree}, so that $ f'_\eq= |0\>\<0|\otimes \rho^{(2)}$, where $\rho^{(2)}$ is the density matrix of the 2-oscillator. Hence, it is clear that $f'_\eq$ is a separable state.

It is interesting to compare the reduced dynamics of the two oscillators system to that of a single oscillator coupled to a bath. In the case of a single oscillator, the reduced dynamics too has the Kossakowski-Lindblad form \cite{Lindblad,Kossa76}, although it has a different set of $L'_i$ operators, $L'_0=\adg a+1/2$, $L'_+=\adg$, and $ L'_-=a$, where $\adg, a$ are the creation and annihilation operators of the oscillator. When written in terms of super-operators, the reduced dynamics has the SU(1,1) symmetry \cite{Ban92,*Tay07}. This reduced dynamics connects all the number bases in this system, and the equilibrium state is the Gibbs state.

For the system we consider, if we trace out the 2-oscillator state from \Eq{Neigenv}, we find that the 1-oscillator equilibrium state becomes
%%%
\begin{align} \label{rhoinfty1}
        f^{(N)}_{\eq,1}= \frac{(1+\nb_0)^N}{Z_N} \sum_{n_1=0,1,2,\cdots}^N  e^{-n_1\beta \w_0} |n_1;n_1\r \,.
\end{align}
%%%
When $N$ is unrestricted, we recover the Gibbs state. Tracing out the 1-oscillator state will produce a similar expression of $f^{(N)}_{\eq,2}$ as in \Eq{rhoinfty1}, except $|n_1;n_1\r$ is replaced by $|N-n_2;N-n_2\r$, whereas the rest of the coefficients remain unchanged.

%%%%%%%%%%%%%%%%%%%%%%%%%%
%       Sectionn         %
%%%%%%%%%%%%%%%%%%%%%%%%%%
\section{Time evolution of states}
\label{sec-timeevo}

To study the time evolution of the reduced system, we first introduce the interaction picture for the reduced dynamics. We denote the density state in this picture as
%%%
\begin{align}
        \rhot\equiv e^{K_0 t}\rho=e^{iH_0't}\rho e^{-iH_0't}\,.
\end{align}
%%%
By expanding $\rhot=\sum_i \ct_i f_i$ and $\rho=\sum_i c_i f_i$ in terms of the time-independent basis $f_i\equiv f^{(N,\Nt)}_{r;\rt}$, the coefficient $\ct_i$ acquires a phase due to the action of $\exp(iK_0 t)$ on $ f_i$,
%%%
\begin{align} \label{tildec}
    \ct_i= c_i \exp(i\tt_i t)\,,
\end{align}
%%%
in which
%%%
\begin{align} \label{thetai}
        \tt_i&=\half (\ome'_0-\del\ome'_0)(N-\Nt)+\half (\wo-\del\wo)(r-\rt)\no
        &\quad-\tfrac{1}{4}\del\ome'_0(N^2-\Nt^2)+\tfrac{1}{4}\del\ome'_0(r^2-\rt^2)\,.
\end{align}
%%%
Since the phase angle vanishes for the coefficients associated to the probability elements $f^{(N)}_{r}$, we have $\ct_i=c_i$. In this situation, we drop the tilde sign on the coefficients to simplify the notation.

In the interaction picture, the equation of motion becomes
%%%
\begin{align}   \label{intrhot}
\frac{\d}{\d t}\rhot=-\Kdt \rhot\,,
\end{align}
%%%
where the effect of $\Kdt$ on the basis state is
%%%
\begin{align} \label{Kdtfinal}
        &\Kdt f_i=-\half  \gam  \nb_0 (2e^{i\del\ome'_0\nu t} L_+ f_i L_- - L_-L_+ f_i -f_i L_-L_+ ) \no
            & -\half \gam (\nb_0+1) (2 e^{-i\del\ome'_0\nu t}L_- f_i L_+ - L_+L_- f_i - f_i L_+L_-)\,,
\end{align}
%%%
in which $\nu$ is already defined in \Eq{Delr}, see App.~\ref{sec-intpic} for details. The extra phase factors in front of two of the terms $L_+ f_i L_-$ and $L_- f_i L_+$ are due to the action of $L_0^2$ in $H_0'$ of \Eq{H0'}. For basis states that lie in the probability subspace, we have $\nu=0$. In this case, $\Kdt f_i$ reduces to $K_d f_i$.

By noting that the $f^{(N,\Nt)}$ and $f^{(\Nt,N)}$ subspaces are related by the following relations, $f^{(\Nt,N)}_{\rt;r}=[f^{(N,\Nt)}_{r;\rt}]^\dg$, $K_0 f^{(\Nt,N)}_{\rt;r}=[K_0f^{(N,\Nt)}_{r;\rt}]^\dg$, and $\Kdt f^{(\Nt,N)}_{\rt;r}=[\Kdt f^{(N,\Nt)}_{r;\rt}]^\dg$, we obtain
%%%
\begin{align} \label{Kfadj}
        K f^{(\Nt,N)}_{\rt;r}=[K f^{(N,\Nt)}_{r;\rt}]^\dg\,.
\end{align}
%%%
Hence, we can deduce the action of $K$ on one subspace from the other. We will next illustrate the general features of the time evolution of the system with a few examples.

%%%%%%%%%%%%%%%%%%%%%%%%%%
%       Sectionn         %
%%%%%%%%%%%%%%%%%%%%%%%%%%
\subsection{States in lower subspaces}

We begin by studying the time evolution of a state dwelling in the subspaces up to $N,\Nt=1$. The density matrix is
%%%
\begin{align} \label{rho12}
\rhot(t)&= d(t) f^{(0)}_{0}+\left[ \gt(t) f^{(0,1)}_{0;1}+ \htt(t) f^{(0,1)}_{0;-1}+\text{h.c.}\right] \no
        &\quad+ a(t) f^{(1)}_{1}+b(t) f^{(-1)}_{-1}+\left[ \ct(t) f^{(1,0)}_{1;0} +\text{h.c.}\right]\,,
\end{align}
%%%
where the coefficients are subjected to the normalization condition $d(t)+\at(t)+\btil(t)=1$ and the positivity conditions of $\rhot$. The effect $\Kdt f_i$ can be worked out using \Eqs{Kdtfinal}, \eqref{LNr+} and \eqref{LNr-}. By equating the coefficients associated to the same basis state $f_i$ on both sides of \Eq{intrhot}, we find that the coefficients evolve as
%%%
\begin{subequations}
\begin{align} \label{dt}
    \dot{d}&=0\,,\\
    \dot{\at}&=  -\gam(1+\nb_0) \at+\gam\nb_0 \btil\,, \label{a1b1c1t}\\
    \dot{\btil}&= \gam(1+\nb_0) \at-\gam\nb_0 \btil\,,\label{b1c1t}\\
    \dot{\ct}&= -\gam(\nb_0+\half) \ct\,. \label{Kf11}\\
        \dot{\gt}&=-\half\gam(\nb_0+1)\gt\,,\label{gt}\\
        \dot{\htt}&=-\half\gam\nb_0 \htt\,,\label{htt}
\end{align}
\end{subequations}
%%%
where we have omitted the time dependence on the coefficients for simplicity.

We first make a few observations. We find that only the coefficients under the same $f^{(N,\Nt)}$ subspace are connected. We also find that the coefficients for the $f^{(1,1)}$ subspace, namely, $a, \btil, \ct$, evolve in exactly the same way as the amplitude damping channel for qubits \cite{Nielsen}. In general, it can be shown that the $f^{(N,N)}$ subspace evolves similar to the $N$-level system under the cascade process with a single decay constant in vacuum, $\gam$,
%%%
\begin{align} \label{cascade}
       |N\> \leftrightarrow |N-1\> \leftrightarrow \cdots \leftrightarrow |1\>   \leftrightarrow|0\>\,,
\end{align}
%%%
as depicted in \Fig{fig1}. For instance, the dynamics in the $f^{(2,2)}$ subspace behaves similar to the three-level system \cite{Hioe82} under the cascade process.

The solutions to \Eqs{dt}-\eqref{htt} are
%%%
\begin{subequations}
\begin{align}
        d(t)&=d_0\,,\\
        \at(t)&=\at_0 e^{-\gam(1+2\nb_0)t}+\frac{\nb_0(1-d_0)}{1+2\nb_0} \left[1-e^{-\gam(1+2\nb_0)t}\right]\,,\label{aN12}\\
        \btil(t)&=1-d_0-\at(t)\,,\\
         \ct(t)&= \ct_0 e^{-(2\nb_0+1)\gam t/2}\,,\label{ct}\\
        \gt(t)&=\gt_0 e^{-\gam(\nb_0+1)t/2}\,,\label{gt0}\\
        \htt(t)&=\htt_0 e^{-\gam\nb_0 t/2}\,,\label{gt0b}
\end{align}
\end{subequations}
%%%
where $d_0$ denotes the value of $d$ at $t=0$, and etc. From these expressions, we learn that the diagonal components eventually settle down at some equilibrium values, whereas all the off-diagonal components vanish asymptotically. In the general $f^{(N,N)}$ subspace, we find that the diagonal coefficients contain the time exponential factor $\exp[-N\gam(2\nb_0+1)t]$, hence they evolve towards the equilibrium value more rapidly, whereas all the off-diagonal coefficients vanish asymptotically. However, as already shown in Sec.~\ref{sec-EigProb}, for the special case of zero temperature, the lowest off-diagonal coefficient in each subspace does not experience decoherence, as is clear from the expression of $\htt(t)$ in \Eq{gt0b}, which is independent of time when $\nb_0=0$.

%%%%%%%%%%%%%%%%%%%%%%%%%%
%       Sectionn         %
%%%%%%%%%%%%%%%%%%%%%%%%%%
\subsection{States involving infinite number of subspaces}
\label{sec-GenSt}

We now investigate the time evolution of states involving the general subspaces. In principle, the initial states can be decomposed into basis states in the various $f^{(N,\Nt)}$ subspaces, and the time evolution can be analyzed subsequently. Since this is a tedious and not an illuminating process, we will discuss the general features of the time evolution by comparing the initial and the equilibrium states with two examples. To simplify the expressions, we will make use of the dimensionless position coordinate
\begin{align}   \label{xdless}
    x_i\equiv \sqrt{\frac{m_i \w_i}{\hbar}} \,q_i\,,
\end{align}
%%%
where $m_i$ is the mass of the $i$-oscillator and $q_i$ is the ordinary position coordinate with the dimension of length.

\begin{figure}[tb]
\centering
\includegraphics[width=3.2in]{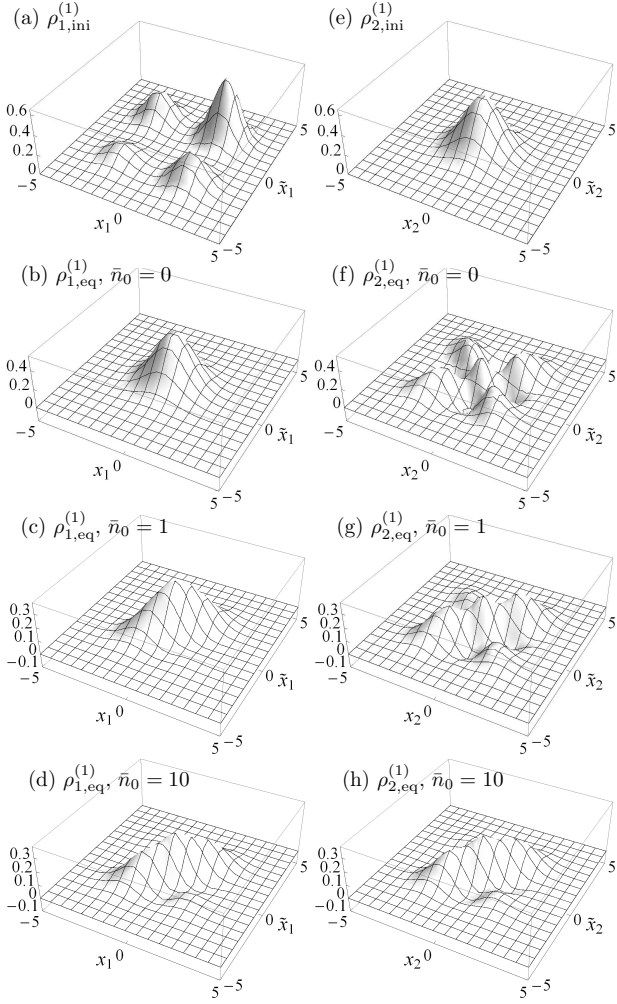}
\caption{Initial and equilibrium configurations of the individual modes of example (1) for several temperatures, with parameters $a=2$ and $N_1/N_2=2$ in \Eq{2peak}.}
\label{fig2}
\end{figure}

(1) Assume that the 1-oscillator is initially a superposition of two Gaussian states centered at $x_1=\pm a$, respectively \cite{Giulini},
%%%
\begin{align}   \label{2peak}
    \phi_1(x_1)&=N_1 e^{-(x_1-a)^2}+N_2 e^{-(x_1+a)^2}= \sum_{n=0}^\infty c_n \<x_1|n\> \,,
\end{align}
%%%
where $N_1, N_2$ are the normalization constants that give the relative height between the two Gaussians. In \Eq{2peak}, we decompose $\phi_1$ in terms of the harmonic oscillator wave function $\<x|n\>=H_n(x) \exp(-x^2/2)/\sqrt{2^n n! \sqrt{\pi}}$, with the expansion coefficient $c_n=\<n|\phi_1\>$, and $H_n(x)$ is the Hermite polynomial. As an example, we choose $a=2$ and $N_1/N_2=2$. A little calculation shows that $c_2$ is the dominant term,
%%%
\begin{align}   \label{cn}
    \{c_0,c_1,c_2,\cdots\}=\{0.34, 0.22, 0.78, 0.23, 0.41, 0.05, \cdots\}\,.
\end{align}
%%%
The density matrix $\rho^{(1)}_{1,\ini}(x_1,\xt_1)=\<x_1|\phi_1\>\<\phi_1|\xt_1\>$ in the coordinate space is plotted in \Fig{fig2}(a).

%%%
%%%

Consider an initially uncorrelated composite system of two oscillators, $\Phi^{(1)}_\ini$, with the 2-oscillator initially in the ground state, $\rho^{(1)}_{2,\ini}=|0\>\<0|$,
%%%
\begin{align}   \label{Phi1}
    \Phi^{(1)}_\ini\equiv\rho^{(1)}_{1,\ini} \otimes \rho^{(1)}_{2,\ini}&=\sum_{n,m=0}^\infty c_nc^*_m |n,0\>\<m,0|\no
    &=\sum_{n,m=0}^\infty c_nc^*_m f^{(n,m)}_{n;m}\,.
\end{align}
%%%
The spatial profile of $\rho^{(1)}_{2,\ini}(x_1,\xt_2)$ is plotted in \Fig{fig2}(e). Under the reduced dynamics, the off-diagonal coefficients undergo exponential decay, and $\Phi^{(1)}_\ini$ eventually evolves into the equilibrium state
%%%
\begin{subequations}
\begin{align}   \label{Phi1eq}
      \Phi^{(1)}_\eq &= \sum_{N=0}^\infty |c_N|^2 f^{(N)}_\eq \\
      &=\sum_{N=0}^\infty |c_N|^2 \!\!\sum_{n_2=0}^N \frac{p_{N-2n_2}}{Z_N} |N-n_2,n_2\>\<N-n_2,n_2|\,,
\end{align}
\end{subequations}
%%%
cf.~\Eq{Neigenv} for the expression of $f^{(N)}_\eq$.

In the special case of zero temperature we have $\nb_0=0$. The 1-oscillator then settles down to the ground state $\rho^{(1)}_{1,\eq}\equiv \tr_2 \Phi^{(1)}_\eq=|0\>\<0|$ with a Gaussian profile as depicted in \Fig{fig2}(b). Notice that the off-diagonal peaks in \Fig{fig2}(a) have decohered away, much like its single oscillator counterpart in a thermal bath \cite{Giulini}. The information carried by the 1-oscillator in the $c_i$s is inherited by the 2-oscillator to some extent, as can be seen in the density matrix of the 2-oscillator,
%%%
\begin{align}   \label{Phi1rho2}
    \rho^{(1)}_{2,\eq} \equiv \tr_1 \Phi^{(1)}_\eq&=\sum_{N=0}^\infty |c_N|^2 |N\>\<N| \,,
\end{align}
%%%
in which only the highest energy level in each of the 2-oscillator subspaces is occupied. Since $c_2$ is the dominant term, $\rho^{(1)}_{2,\eq} \approx |c_2|^2 |2\>\<2|$. The plot in the coordinate space then gives a characteristic three-peak profile of the wave function $\<x|2\>$ along the diagonal, see \Fig{fig2}(f).

When temperature increases, the populations of the 1- and 2-oscillator start to distribute accordingly among different levels in the $f^{(N)}$ subspace. The reduced states are then approximately given by
%%%
\begin{align}   \label{chi1Tneq1}
    \rho^{(1)}_{1,\eq}\approx \frac{|c_2|^2}{Z_2} \big( p^{(2)}_2 |2\>\<2| +p^{(2)}_0 |1\>\<1|+p^{(2)}_{-2}|0\>\<0|\big)\,,\\
    \rho^{(1)}_{2,\eq}\approx \frac{|c_2|^2}{Z_2}  \big( p^{(2)}_{-2} |2\>\<2| +p^{(2)}_0 |1\>\<1|+p^{(2)}_{2}|0\>\<0|\big)\,.\label{chi1Tneq2}
\end{align}
%%%
As a result, the three-peak profile is smoothed out, as shown in \Figs{fig2}(c) and \ref{fig2}(g) for $\nb_0=1$. By comparing \Eq{chi1Tneq1} with \eqref{chi1Tneq2}, we notice that the order of the level populations between the 1- and 2-oscillator, i.e., $p^{(2)}_2, p^{(2)}_0$ and $p^{(2)}_{-2}$, is reversed. This feature recurs in the other subspaces as well. For large temperature, $p^{(N)}_{i,\eq}\approx \nb_0^N$ for all $i$. Consequently, we obtain a uniform distribution among all the levels. In this limit, $\rho^{(1)}_{1,\eq}\approx \rho^{(1)}_{2,\eq}$, as can be seen by comparing \Figs{fig2}(d) and \ref{fig2}(h) for $\nb_0=10$.

(2) We next consider an initially entangled state,
%%%
\begin{align}   \label{Phi2in}
    |\Phi^{(2)}_\ini\>=\sum_{n=0}^\infty c_n |n,n\>\,.
\end{align}
%%%
For a comparison with the first example, we choose $c_n$ to be the same as those in \Eq{cn}. In this example, both the oscillators are initially in the same state,
%%%
\begin{align}   \label{chi2in}
    \rho^{(2)}_{1,\ini}=\rho^{(2)}_{2,\ini}=\sum_{n=0}^\infty |c_n|^2 |n\>\<n|\approx |c_2|^2 |2\>\<2|\,.
\end{align}
%%%
The density matrix of the system, $\Phi^{(2)}_\ini \equiv|\Phi^{(2)}_\ini\>\<\Phi^{(2)}_\ini|$, evolves into the equilibrium state
%%%
\begin{align}   \label{Phi2eq}
    \Phi^{(2)}_\eq &= \sum_{N=0}^\infty |c_N|^2 f^{(2N)}_\eq \,,
\end{align}
%%%
which has been shown to be separable in Sec.~\ref{sec-EigProb}. Hence, the entanglement between the pair of oscillators are lost eventually. In fact, any initial entanglement in the reduced system vanishes asymptotically in view of the separability of the equilibrium state $f^{(N)}_\eq$ \eqref{Neigenv}.

A comparison of \Eq{Phi2eq} with \Eq{Phi1eq} also shows that the choice of the superpositions $|n,n\>$ in \Eq{Phi2in} have resulted in the exclusion of the odd $f^{(2N+1)}$ subspace from the equilibrium state. This is a consequence of the fact that the set of odd number subspaces are absent from the initial state. Therefore, by specifically preparing the initial state, some subspaces could be excluded from the equilibrium state. In this model, states with different initial conditions may evolve into different classes of equilibrium states.

%%%%%%%%%%%%%%%%%%%%%%%%%%
%       Sectionn         %
%%%%%%%%%%%%%%%%%%%%%%%%%%
\section{Conclusion}
\label{sec-conclusion}

We have considered a system of two oscillators collectively coupled to a field through a three-body interaction. The model is applicable within a time frame in which nonlinear interaction is dominant. The two oscillator modes become effectively coupled as a result of their collective interaction with the field. Consequently, the reduced dynamics possesses the SU(2) symmetry that is common to finite-level systems, which leads to non-Gaussian equilibrium states for the collective modes. These are metastable states until linear interactions between the individual oscillators and the field takes over and drives them to new equilibrium states. The results suggest that new forms of equilibrium states could emerge when subsystems are collectively coupled to the environment under different symmetry of the reduced dynamcis.

It is interesting to further explore the implications and manifestations of the results in other systems, such as in the orbital motion of two-electron quantum dots \cite{Chizhov12} and light-phonon systems \cite{Chizhov91}. In the later case, the existence of metastable states of photons may prevent photons from thermalizing too rapidly with phonon bath. This may have interesting implications in photosynthetic systems, where the role of phonon is played by the vibrational modes of photosynthetic reaction centers \cite{Plenio08}. We leave these interesting investigations to future works.

%%%%%%%%%%%%%%%%%%%%%%%%%%
%    Acknowledgmentss    %
%%%%%%%%%%%%%%%%%%%%%%%%%%
\acknowledgments

We thank Dr.~G.~Ord\'o\~nez for very helpful comments, suggestions and discussions, and for carefully reading the manuscripts. We also thank the referees for their comments and suggestions. Support by the Fundamental Research Grant Scheme (FRGS), Grant No.~FP009-2011A, under the Malaysian Ministry of Higher Education (MOHE) is gratefully acknowledged.

\newpage

\appendix

%%%%%%%%%%%%%%%%%%%%%%%%%%
%       Sectionn         %
%%%%%%%%%%%%%%%%%%%%%%%%%%
\section{Oscillators coupled by SU(2) coupling interactions}
\label{sec-SU2}

Using the operator \cite{SU2AnisoOscLin12,SU2Chen11}
%%%
\begin{align}   \label{U}
U=\exp(-i\phi L_0)\exp(-i\theta L_2)\,,
 \end{align}
%%%
where $L_i$ are the generators of the SU(2) group \eqref{L0}-\eqref{L12}, the $a_i$s are related to a corresponding set of $b_i$ operators by
%%%
\begin{align}   \label{Uaa'}
   & \left(\begin{array}{c}
                    a_1\\
                    a_2\\
            \end{array} \right)
            =U \left(\begin{array}{c}
                    b_1\\
                    b_2\\
            \end{array} \right) U^\dg\no
            &\quad=            \left(\begin{array}{cc}
                   e^{i\phi/2}\cos(\theta/2)& e^{-i\phi/2}\sin(\theta/2)\\
                   -e^{i\phi/2}\sin(\theta/2)& e^{-i\phi/2}\cos(\theta/2)\\
            \end{array} \right)\left(\begin{array}{c}
                    b_1\\
                    b_2\\
            \end{array} \right)\,.
\end{align}
%%%
The family of Hamiltonian related to the oscillators subsystem is then given by
%%%
\begin{align}   \label{H12}
    H_{12} &= \w_1 {a_1}^\dg a_1 + \w_2 {a_2}^\dg a_2 \no
    &= \frac{\wo'}{2} ({b_1}^\dg b_1 + {b_2}^\dg b_2) +\wo \cos\theta L_0\no
    &\qquad + \wo \sin \theta (\cos \phi L_1 + \sin\phi L_2)\,,
\end{align}
%%%
with the corresponding interaction
%%%
\begin{align} \label{HVb}
    V&\sim \adg_1 a_2 a_k +a_1 \adg_2 \adg_k \no
        &= \big[-\sin\theta L_0 +\cos \theta (\cos \phi L_1 +\sin \phi L_2)\big] (a_k+\adg_k)\no
        &\qquad -i(\sin\phi L_1 -\cos\phi L_2) (a_k-\adg_k)\,,
\end{align}
%%%
where $a_1, a_2$ in $L_i$ \eqref{L0}-\eqref{L12} are replaced by $b_1, b_2$, respectively. The second equality is obtained by substituting \Eq{Uaa'} into the first line of \Eq{HVb}. Note that $N$ \eqref{N} also commutes with $U$. Hence, the total occupation quantum number of the oscillators remain the same under the change of basis. Note also that the vacuum state does not change under $U$, i.e., $|0,0\>'=U^\dg |0,0\>=|0,0\>$.

%%%%%%%%%%%%%%%%%%%%%%%%%%
%       Sectionn         %
%%%%%%%%%%%%%%%%%%%%%%%%%%
\section{Comparison with anharmonic interactions}
\label{sec-anharmonic}

The anharmonic interactions couple the position operators of the three species. Using dimensionless position coordinate defined in \Eq{xdless}, the position operator $\xh_i$ is related to the $a_i$ by $\xh_i=(a_i+\adg_i)/\sqrt{2}$. The anharmonic interactions then take the form
%%%
\begin{align} \label{Vall}
  V'=\lam \sum_k \frac{v(\w_k)}{\sqrt{\Omega/2\pi}} (a_1+\adg_1)(a_2+\adg_2)(a_k+ \adg_k)\,.
\end{align}
%%%
In this situation, we find that aside from the $\w_0$ resonant mode, there is a second resonant mode that occurs at a higher frequency $\w'_0$ compared to $\w_0$,
%%%
\begin{align}  \label{w12}
        \wo'&= \w_1+\w_2\,.
\end{align}
%%%
This mode is initiated by the interaction terms $a_1 a_2 \adg_k +\text{h.c.}$, where $\text{h.c.}$ denotes hermitian conjugate. If the occupation number of the $\w'_0$ mode in the field is small enough then this mode is not excited. This could be achieved if the temperature of the bath, $T$, is low enough so that the condition
%%%
\begin{align}  \label{wT}
        \w_0 < k_B T \ll \w'_0
\end{align}
%%%
is satisfied, where $k_B$ is the Boltzmann constant. Another way to achieve this is to formally impose a high frequency cut-off $\w_c$ to the form factor $v(\w_k)$ in the interaction \eqref{Vall} so that
%%%
\begin{align}  \label{w'0T}
        k_B T<\w_c \ll \w'_0
\end{align}
%%%
is satisfied. Both conditions are separately consistent with the $\lam^2t$-approximation \cite{VanHove57,Prigogine62}, or the Born-Markov approximation \cite{Louisell72,Breuer}, used to derive the Markovian master equation of the system. Condition \eqref{wT} implies that $\w_1, \w_2$ should be of the same order, since $\w_1-\w_2<k_B T$ for small $T$, whereas condition \eqref{w'0T} implies that they should not be too small, since we require $\w_c\ll \w_1+\w_2$ for a large cut-off frequency. In the latter situation, since $\w_1$ and $\w_2$ are not small, only a few lower energy modes will be excited in the dynamics, see \Fig{fig1} for the energy levels of the system.

Aside from the resonant modes, there are two other virtual modes that involve the field quanta with negative frequencies, i.e., $\w'_v=-(\w_1+\w_2)$ and $\w_v=-(\w_1-\w_2)<0$. They are initiated by the interaction terms $\adg_1\adg_2\adg_k+\text{h.c.}$ and $a_1 \adg_2 a_k+\text{h.c.}$ of \Eq{Vall}, respectively. The $\w'_v$ mode is a fast rotating mode so that its contribution to the reduced dynamics averages to zero in the relaxation time scale, $\tau_R\sim 1/\gam$, where the $\lam^2t$-approximation is valid. This is the rotating-wave approximation \cite{Cohen-Tannoudji,Breuer} usually implemented in the Markovian limit.

The virtual mode $\w_v$ does not contribute to the reduced dynamics on the level of the $\lam^2t$-approximation with the collision operator $\psi^{\nuv}_2$ defined in \Eq{psi2nu}. A similar example is provided by Ref.~\onlinecite{Barsegov02} for the reduced dynamics of a single oscillator coupled to a field, in which the interactions contain a virtual transition mode. The effect of a possible extension of the collision operator $\psi^{\nuv}_2$ \eqref{psi2nu} to include the contribution of the virtual transitions is presented in the next appendix.

It is also interesting to note that if instead of assuming $\w_1 >\w_2$ in our discussion so far, the opposite situation $\w_2>\w_1$ would interchange the role played by the $\w_v$ and the $\w_0$ modes, i.e., now $\w_0$ becomes a virtual mode, whereas $\w_v$ becomes a resonant mode. However, the $\w'_0$ and $\w'_v$ modes are not affected by this change.

In summary, in comparison to the full anharmonic interactions, the higher frequency resonant mode $\w'_0$ is not included in the interaction Hamiltonian of the system we consider here. The contribution of this mode is negligible compared to the $\w_0$ mode at low temperature, or when a frequency cut-off is imposed to the form factor below $\w'_0$. Moreover, out of the two virtual modes we have discussed, the effect of the $\w'_v$ mode is effectively dropped under the rotating-wave approximation, whereas the $\w_v$ mode does not affect the dissipative part of the reduced dynamics using the usual definition of the collision operator.

%%%%%%%%%%%%%%%%%%%%%%%%%
%       Sectionn        %
%%%%%%%%%%%%%%%%%%%%%%%%%
\section{Extension of $\psi_2$}
\label{sec-extpsi2}

%%%
\begin{figure}[tb]
\centering
\includegraphics[width=2.4in]{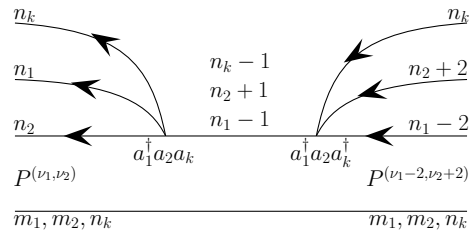}
\caption{$P^{(\nu_1,\nu_2)}\psi_2 P^{(\nu_1-2,\nu_2+2)}$ transition. It is one of the diagrams giving the contribution to the collision operator $\psi_2$ when virtual mode $\w_v$ is considered. The operators on the right vertex cause the virtual transition.}
\label{fig3}
\end{figure}
%%%

Using the usual definition of the collision operator up to the second order in the coupling constant \cite{Petrosky97},
%%%
\begin{align}   \label{psi2nu}
        \psi^{\nuv}_2 &= P^{\nuv} \psi_2 P^{\nuv} \\
        &\equiv P^{\nuv}\L_0P^{\nuv}-\lam^2 P^{\nuv}\L_V Q^{\nuv}\frac{1}{\L_0- \vec{\w}\cdot \nuv -i \eps} Q^{\nuv}\L_V P^{\nuv} \,,\label{psi2}
\end{align}
%%%
the virtual mode $\w_v$ does not contribute to the collision operator on the level of $\lam^2t$-approximation. However, if we incorporate the transitions between different $P^{\nuv}$ subspaces into the definition of the collision operator \eqref{psi2nu}, the virtual mode $\w_v$ may contribute to the dissipative part of the reduced dynamics as follows \cite{[{See also the footnote 5 of }][{}]Tay06}.

We define the collision operator by $\psi'_2\equiv P\psi_2 P$, where $P\equiv \sum_{\nuv} P^{\nuv}$. \Fig{fig3} shows an example of a possible transition of $\psi'_2$ that involves different $P^{\nuv}$ subspaces. As indicated in the figure, the interaction vertex on the right is due to the virtual transition $\adg_1 a_2 \adg_k$, whereas the vertex on the left is due to a real transition mode in the original model \eqref{HV}. A standard calculation shows that \Fig{fig3} contributes a term $(\adg_1 a_2)^2 \fh$ to the dissipative part of the reduced dynamics, in addition to the contributions by the real transitions. When taking into account all these additional transitions, the terms $L_+\fh L_+$, $L_+^2 \fh$, $\fh L_+^2$ and their hermitian conjugates, are added to the dissipative operator, resulting in a not completely positive reduced dynamics \cite{Pechukas94,*Shaji05}.

We note that this extension to the definition of the collision operator does not affect the reduced dynamics of the original model \eqref{HV}, since all the possible transitions in this model involve the same $P^{\nuv}$ subspace. We already learned that in this situation they lead to a completely positive reduced dynamics  \cite{Lindblad,Kossa76} with the Kossakowski-Lindblad form \eqref{Kd}.

%%%%%%%%%%%%%%%%%%%%%%%%%
%       Sectionn        %
%%%%%%%%%%%%%%%%%%%%%%%%%
\section{Rotational symmetry under $L_0$}
\label{sec-L0sym}

In this appendix, we show that $K$ is invariant under a rotation along the $L_0$ axis. Indeed, we find that
%%%
\begin{align} \label{L0Lpm}
        L'_\pm &\equiv e^{i \theta L_0} L_\pm e^{-i\theta L_0}= e^{\pm i\theta} L_\pm \,,
\end{align}
%%%
by using the commutation relations $[L_+, L_-]=2L_0$ and $[L_0, L_\pm]=\pm L_\pm$. Since $L_+$ and $L_-$ appear pairwise in $K$, cf.~\eqref{K}-\eqref{Kd}, the phase in \Eq{L0Lpm} cancels out. Added by the fact that $L_0$ commutes with $N$, we conclude that $K$ is invariant under the rotation.

%%%%%%%%%%%%%%%%%%%%%%%%%
%       Sectionn        %
%%%%%%%%%%%%%%%%%%%%%%%%%
\section{$SU(2)$ generalized coherent states reside in $f^{(N,N)}$ subspace}
\label{sec-SU2cohst}

The SU(2) generalized coherent states \cite{Perelomov72,Wodkiewicz85,Buzek89},
%%%
\begin{align}   \label{SU2gcs}
    |\tau\>_N=(1+|\tau|^2)^{-\frac{N}{2}}\sum_{n_1=0}^N
            \sqrt{
            \bigg(\!\!
            \begin{array}{c}
            N \\ n_1\\
            \end{array}\!\!
            \bigg)}
                    \tau^{n_1} |n_1,N-n_1\>\,,
\end{align}
%%%
reside in the corresponding $f^{(N,N)}$ subspace of the system, where $\tau=\tan(\theta/2)\exp(-i\phi)$, in which $\theta, \phi$ are the parameters that label the coherent states. In terms of the $f^{(N,N)}$ basis, we have
%%%
\begin{subequations}
\begin{align}   \label{SU2DM}
    |\tau\>_N\<\tau|&=\sum_{n,m}^N c^{(N,N)}_{n;m}f^{(N,N)}_{N-2n;N-2m}\,,\\
    c^{(N,N)}_{n,m}&=\frac{1}{2^N} (\sin 2\theta)^N \sqrt{
            \bigg(\!\!
            \begin{array}{c}
            N \\ n
            \end{array}\!\!
            \bigg)
            \bigg(\!\!
            \begin{array}{c}
            N \\ m
            \end{array}\!\!
            \bigg)}
            e^{i\phi(n-m)}\,.
\end{align}
\end{subequations}
%%%

%%%%%%%%%%%%%%%%%%%%%%%%%
%       Sectionn        %
%%%%%%%%%%%%%%%%%%%%%%%%%
\section{$K_d$ in the interaction picture}
\label{sec-intpic}

In this appendix, we calculate the effect of $\Kdt$ on the $f_i$ basis. Since $L_0$ and $N$ commute, we have
%%%
\begin{align} \label{H0'comm}
        e^{iH_0' t}=e^{i(\ome'_0-\del\ome'_0)Nt/2-i\del\ome'_0N^2t/4}e^{i\del\ome'_0L_0^2t}e^{i(\wo -\del\wo) L_0 t} \,.
\end{align}
%%%
Using \Eq{L0Lpm}, we find that
%%%
\begin{align} \label{Lpm'}
        e^{i(\wo -\del\wo) L_0 t}L_\pm e^{-i(\wo -\del\wo) L_0 t}
        =e^{\pm i (\wo -\del\wo)t} L_\pm\,.
\end{align}
%%%
Therefore, $\Kdt=e^{i\del\ome'_0L_0^2t}K_d e^{-i\del\ome'_0L_0^2t}$,
since $L_+$ and $L_-$ appear pairwise in $K_d$, and $N$ commutes with $L_0, L_\pm$.
Furthermore, using
%%%
\begin{align} \label{L02Lpm}
        [L_0^2,L_\pm]=\pm(2L_0\mp 1)L_\pm=\pm L_\pm(2L_0\pm 1)\,,
\end{align}
%%%
we can show that
%%%
\begin{align} \label{Lpmtilde}
        \Lt_\pm\equiv e^{i\del\ome'_0L_0^2t}L_\pm e^{-i\del\ome'_0L_0^2t}
            &=L_\pm e^{\pm i\del\ome'_0(2L_0\pm 1) t}\\
            &=e^{\pm i\del\ome'_0(2L_0\mp 1) t}L_\pm\,,
\end{align}
%%%
where $\Lt_\pm^\dg=\Lt_\mp$. Using the relation \eqref{L0st}, we finally obtain
\Eq{Kdtfinal}. Therefore, $\Kdt$ differs from $K_d$ by a phase factor in two of the terms. For basis vectors that lie in the probability subspace, $\nu\equiv r-\rt=0$, and we get $\Kdt=K_d$.

%\bibliography{BosonSU2}{}

\providecommand{\noopsort}[1]{}\providecommand{\singleletter}[1]{#1}%

\end{document}